\documentclass[a4paper,twoside]{article}
\usepackage{tikz}
\usepackage{epsfig}
\usepackage{subcaption}
\usepackage{calc}
\usepackage{amssymb}
\usepackage{amstext}
\usepackage{amsmath}
\usepackage{amsthm}
\usepackage{multicol}
\usepackage{pslatex}
\usepackage{apalike}
\usepackage{SCITEPRESS}     
\usepackage{balance}
\usepackage{flushend}
\usepackage{booktabs}
\usepackage{hyperref}
\usepackage[utf8]{inputenc}

\hypersetup{
    colorlinks=true,
    linkcolor=black,
    filecolor=magenta,      
    urlcolor=magenta,
    citecolor=black,
}

\begin{document}

\title{Ambient Lighting Generation for Flash Images with Guided Conditional Adversarial Networks}

\author{\authorname{José Chávez\sup{1}, Rensso Mora\sup{1} and Edward Cayllahua-Cahuina\sup{2}}
\affiliation{\sup{1}Department of Computer Science, Universidad Católica San Pablo, Arequipa, Perú}
\affiliation{\sup{2}LIGM, Université Paris-Est, Champs-sur-Marne, France}
\email{\{jose.chavez.alvarez, rvhmora\}@ucsp.edu.pe, edward.cayllahua@esiee.fr}
}




\abstract{To cope with the challenges that low light conditions produce in images, photographers tend to use the light provided by the camera flash to get better illumination.~Nevertheless, harsh shadows and non-uniform illumination can arise from using a camera flash, especially in low light conditions.~Previous studies have focused on normalizing the lighting on flash images; however, to the best of our knowledge, no prior studies have examined the sideways shadows removal, reconstruction of overexposed areas, and the generation of synthetic ambient shadows or natural tone of scene objects.~To provide more natural illumination on flash images and ensure high-frequency details, we propose a generative adversarial network in a guided conditional mode.~We show that this approach not only generates natural illumination but also attenuates harsh shadows, simultaneously generating synthetic ambient shadows.~Our approach achieves promising results on a custom FAID dataset, outperforming our baseline studies.~We also analyze the components of our proposal and how they affect the overall performance and discuss the opportunities for future work.}

\onecolumn \maketitle \normalsize \setcounter{footnote}{0} \vfill

\section{\uppercase{Introduction}}
\label{sec:introduction}

\begin{figure}[!ht]
\centering
    \begin{subfigure}[t]{0.235\textwidth}
      \centering
      \includegraphics[width=\linewidth]{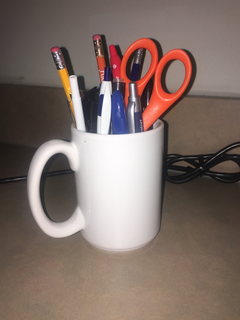}
      \subcaption{Flash image} \label{fig:fig1a}
    \end{subfigure}\hfill
    \begin{subfigure}[t]{0.235\textwidth}
      \centering
      \includegraphics[width=\linewidth]{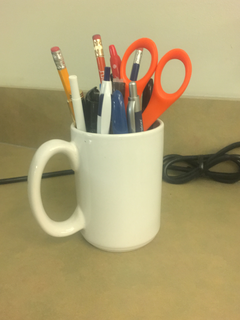}
      \subcaption{Ambient image} \label{fig:fig1b}
    \end{subfigure}
    \vspace{8pt}
\caption{A comparison of a flash image and an ambient image.~(a) Image with camera flash illumination.~The image suffers from harsh shadows, dark areas and bright areas.~(b) Image with available ambient illumination.~In this image, the illumination is more uniform, natural, and the image has not sideways shadows.~Images extracted from FAID~\cite{aksoy2018dataset}.}
\label{fig:fig1}
\end{figure}

\noindent Scenes with low light conditions are challenging in photography, cameras usually produce noisy and/or blurry images.~In these situations, people usually use an external device such as a camera flash, thus, creating flash images.~However, when the light from the flash is pointing directly at the object, the light can be too harsh for the scene and create a non-uniform illumination.~Comparing a flash image with its respective image with ambient illumination, it is clear that the illumination is more natural and uniform because the available light can be more evenly distributed (see Figure~\ref{fig:fig1}).

Researchers have studied the enhancement of flash images~\cite{Petschnigg:2004:DPF:1015706.1015777,Eisemann:2004:FPE:1015706.1015778,Agrawal:2005:RPA:1073204.1073269,CAPECE201928}, producing enhanced images by the combination of such ambient and flash images, or normalizing the illumination on flash image in a controlled environment (backdrop and studio lighting), but without replicating the natural skin tone of people.~However, in a real scenario with low light conditions, there is no information about how the ambient image is.~On the other hand, on scenarios without a backdrop, objects away from the camera will have very low illumination, thus, creating dark areas in the image, considering that there is only the illumination of the camera flash.~Consequently, in a real scenario with low light conditions, creating ambient images from flash images poses a very challenging problem.

Prior works handle the enhancement of low light images, where a scene is underexposed; however, on flash images, objects close to the camera tend to be bright and these techniques overexpose these regions.~Our method attenuates the illumination that is close to the camera, and illuminates the underexposed regions at the same time.~Since flash and ambient images represent the same scene, researchers~\cite{CAPECE201928} study the lighting normalization on a flash image by learning the relationship between both images to estimate a relationship between these pair of images, which is added to the respective flash image in a next step, thus, normalizing the illumination on flash images but maintaining high-frequency information.~This approach is not effective to restore overexposed areas due to this region still needs to compute the final result.

In this article, we propose a conditional adversarial network in a guided mode, which follows two objective functions.~First, the reconstruction loss generates uniform illumination and synthetic ambient shadows.~Second, the adversarial loss, which represents the objective function of GANs~\cite{goodfellow2014generative}, forces to model high-frequency details on the output image, and perform a more natural illumination.~Both loss functions are guided through the attention mechanism, which is performed by attention maps based on the input image and ground truth.~The attention mechanism allows to the model to be more robust to overexposed areas and sideways shadows presented on flash images.~It also improves the robustness of the model on inconsistent scene match between pairs of flash and ambient images since they are both usually not perfectly aligned at the moment of capture.~We compare against state-of-the-art enhancement techniques for low light images~\cite{Fu_2016_CVPR,LIME}, and flash images~\cite{CAPECE201928}.~Ablation studies are also performed on the architecture.

Then, the major contributions of this article are:
\begin{itemize}
    \item An attention mechanism to guide a conditional adversarial network on the task of translating from flash images to ambient images.~Giving robustness against overexposed areas and shadows presented on flash and ambient images, and the misaligned scene between both images.~This mechanism guides the adversarial loss to avoid blurry results on regions by discriminating these cases.
    
    \item Our proposed attention mechanism also guides the reconstruction loss to be robust against high-frequency details thought the texture information that the attention map gives.
\end{itemize}

\section{\uppercase{Related Work}}
\label{sec:relatedwork}

\subsection{Low Light Image Enhancement}
\noindent Prior works \cite{Petschnigg:2004:DPF:1015706.1015777,Eisemann:2004:FPE:1015706.1015778,Agrawal:2005:RPA:1073204.1073269} combine the advantages of both ambient and flash images.~These image processing techniques use the information of the image with the available illumination (ambient image) and the image with light from the camera flash (flash image) and create an enhanced image based on both images.~In contrast with these techniques, our model enhances the flash image but without any kind of information of the ambient image.\\

In SRIE~\cite{Fu_2016_CVPR}, the reflectance and illumination are estimated by a weighted variational model, then, the images are enhanced with the reflectance and illumination components.~LIME~\cite{LIME}, on the other hand, enhance the images by the estimation of their illumination maps.~More specific, the illumination map of each pixel is first estimated individually by finding the maximum value in the R, G and B channels, then the illumination map is refined by imposing a structure prior.~This refined illumination map has smoothness texture details.~Both methods SRIE and LIME do not contemplate sideways shadows removal, reconstruction of overexposed areas or generation of synthetic ambient shadows.\\

\subsection{Image-to-Image Translation}

Prior works use symmetric encoder-decoder networks~\cite{ronneberger2015u,Isola_2017_CVPR,Chen_2018_CVPR} for image-to-image translation such as: image segmentation, synthesizing photos, enhancing low light images, etc.~These networks are composed of various convolutional layers, where the input is encoded to a latent space representation and then decoded to estimate the desired output.~Inspired on the U-Net architecture~\cite{ronneberger2015u}, our model employs skip connections to share information between encoder and decoder, to recover spatial information lost by downsampling operations.

In~\cite{CAPECE201928}, a deep learning model turns a smartphone flash selfie into a studio portrait.~The model generates a uniform illumination, but not reproduce the same skin tone of the person under studio lighting.~The encoder part of the network represents the first 13 convolutional blocks of the VGG-16~\cite{Simonyan15}, and the weights of the encoder are initialized with a pre-trained model for face-recognition~\cite{Parkhi15}.~\cite{CAPECE201928} implemented a pre-processing step before feed the inputs and targets to their architecture, each pair of images are filtered.~After this step the encoder-decoder architecture is trained to learn a low-frequency relationship between the flash and the ambient image.~This extra step slows down the computation in the training and testing.~We exploit the transfer learning approach of this model, but we propose an end-to-end architecture where the encoder path is initialized with the VGG-16 pre-trained on the ImageNet dataset~\cite{deng2009imagenet}, thus, making our model for general scenes, not only for faces and without any additional pre-processing step.

\subsection{Conditional GANs}

Conditional GANs~\cite{mirza2014conditional} have been proposed as a general purpose for image-to-image translation~\cite{Isola_2017_CVPR}.~A cGAN is composed of two architectures, the generator, and the discriminator.~Both architectures are fully convolutional networks~\cite{fullycnn}.~On the generator, which represents an encoder-decoder network, each step of the encoder and decoder is mainly composed by convolutional layers.~The generator $G$ and discriminator $D$ are conditioned on some type of information such as images, labels, texts, etc.~In our case, this information represents the flash images $I_f$, and our cGAN learns to map from flash images $I_f$ to ambient images $I_a$.~Thus, the generator synthesizes ambient images $\hat{I}_a$, which can not be distinguished from the real ambient images $I_a$, while the discriminator is trained in adversarial form respect to the generator to distinguish between $I_a$ and $\hat{I}_a$.~As it is shown in pix2pix model~\cite{Isola_2017_CVPR}, this min-max game ensures the learning of high-frequency details unlike using only a reconstruction loss such as the L1 distance or L2 distance, which outputs blur results.

\section{\uppercase{Proposed Method}}
\label{sec:proposal}

\begin{figure}[!ht]
\centering
\includegraphics[width=\linewidth]{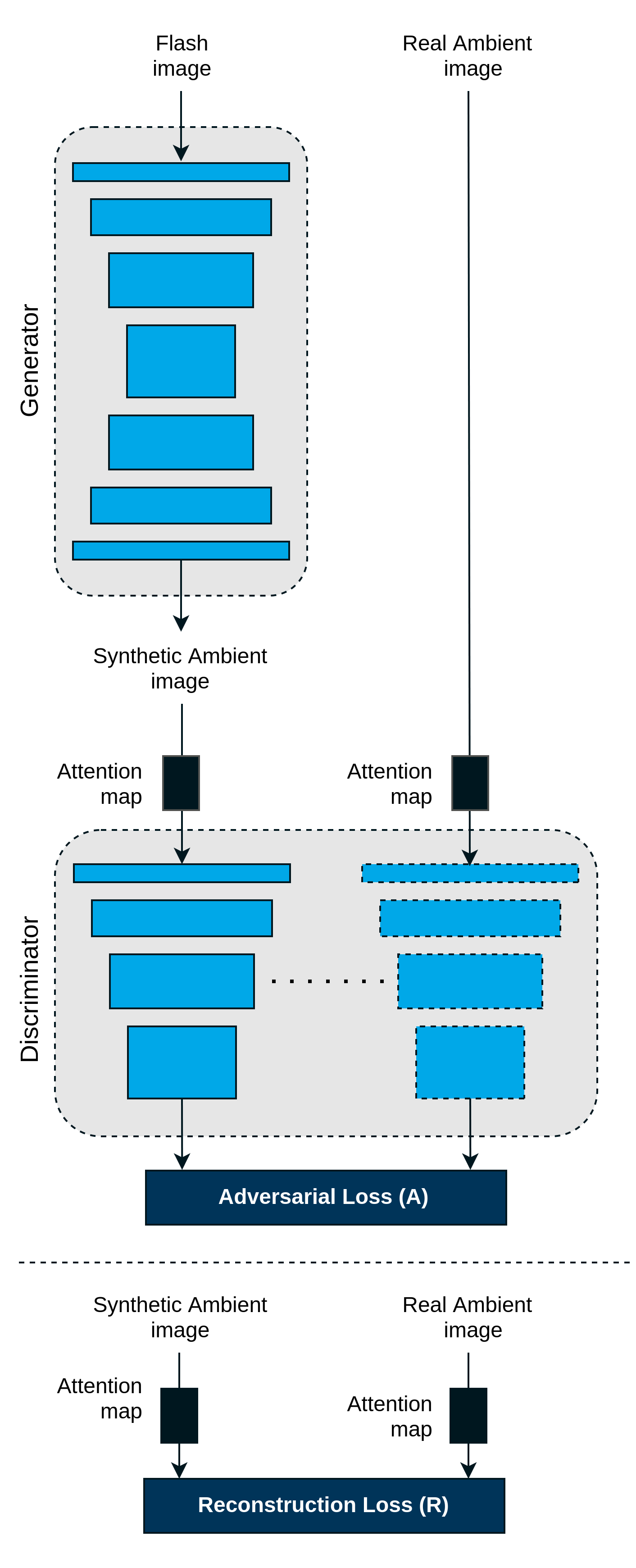}
\caption{Network architecture.~The generator has as its input the flash image $I_f$ and as its output the synthetic ambient image $\hat{I}_a$.~The discriminator network learns through the adversarial loss $\mathbf{A}$ to classify between the real ambient image $I_a$, this is the ambient image that belongs to the training set, and the synthetic ambient image $\hat{I}_a$.~We also set the reconstruction loss $\mathbf{R}$ between $I_a$ and $\hat{I}_a$.~All attention maps are compute thought $I_f$ and $I_a$.}
\label{fig:fig2}
\end{figure}

\noindent Our model is composed of two architectures, generator $G$, and discriminator $D$; and translate from flash images  $I_{f}$ to ambient images $I_{a}$.~Then, the training procedure follows two objectives: the reconstruction loss $\mathbf{R}$, which aims to minimize the distance between the input image~($I_{f}$) and the target image~($I_{a}$); and the adversarial loss $\mathbf{A}$; which represent the objective of the cGAN~\cite{Isola_2017_CVPR}.~Figure~\ref{fig:fig2} illustrates an overall of our architecture model.\\

Both the reconstruction loss $\mathbf{R}$ and the adversarial loss $\mathbf{A}$ are guided by our attention mechanism to ensure a better learning procedure.~The attention mechanism is performed on the entries of $\mathbf{R}$ and $\mathbf{A}$, that is, the ambient image $I_a$ and synthetic ambient image $\hat{I}_a$ first pass through the attention map before the computation of $\mathbf{R}$ and $\mathbf{A}$.

\subsection{Attention Mechanism}

The attention mechanism that we propose aims to guide the reconstruction and adversarial loss.~The mechanism is simple but efficient, we guide both $\mathbf{R}$ and $\mathbf{A}$ with an attention map base on the flash image $I_{f}$ and the ambient image $I_{a}$.~We define the attention map $\mathcal{M}$ as:

\begin{equation}\label{eq:attmap}
\begin{aligned}
\mathcal{M}(i,j)&=1-\frac{1}{C}\sum^{C}_{k=1}\mid I_{a}(i,j,k)-I_{f}(i,j,k)\mid.
\end{aligned}
\end{equation}

In Equation~\ref{eq:attmap}, $C$ represents the number of channels and $\mathcal{M}(i,j)$ the value of the attention map at the position $(i,j)$.~$I(i,j,k)$ represent the pixel value at $(i,j)$ and channel $k$.~Then, $I_{a}$ and $\hat{I}_{a}$ pass though the attention map before compute the reconstruction loss\footnote{we denote $I_{f}\sim p_{\mathbf{data}}\triangleq I_{f}\sim p_{\mathbf{data}}(I_{f})$ and $I_{a}\sim p_{\mathbf{data}}\triangleq I_{a}\sim p_{\mathbf{data}}(I_{a})$ for simplicity.} $\mathbf{R}$ and the adversarial loss $\mathbf{A}$,

\begin{equation}\label{eq:attmapopt}
\begin{aligned} 
I_{a}:=&I_{a}\otimes \mathcal{M},\hat{I}_{a}:=\hat{I}_{a}\otimes \mathcal{M}.
\end{aligned}
\end{equation}

The operation $\otimes$ represents the element-wise multiplication.~Equation~\ref{eq:attmapopt} guides $\mathbf{A}$ and $\mathbf{R}$ to a better learning procedure through the discrimination of overexposed areas, shadows, and scene misalignment, between $I_f$ and $I_a$.~Then $\mathbf{R}$, which represent the L1 distance, and $\mathbf{A}$ are defined as:

\begin{equation}\label{eq:recadvloss}
\begin{aligned} 
\mathbf{R}(G)&=\mathbb{E}_{I_{f}\sim p_{\mathbf{data}}, I_{a}\sim p_{\mathbf{data}}}\left[\left\lVert I_{a} - G(I_{f}) \right\rVert_{1}\right]\\
\mathbf{A}(D,G) &= \mathbb{E}_{I_{a}\sim p_{\mathbf{data}}}\left[\log D(I_{a})\right]\\
                &+ \mathbb{E}_{I_{f}\sim p_{\mathbf{data}}}\left[\log (1-D(G(I_f)))\right].
\end{aligned}
\end{equation}

By this operation, the reconstruction loss $\mathbf{R}$ is conducted to learn the normalization of the lighting, discriminating the high-frequency details because the attention map $\mathcal{M}$ gives this information by the element-wise multiplication.~$\mathcal{M}$ also guides $\mathbf{R}$ to be robust for the misaligned scene between flash and ambient images.~On the other hand, the adversarial loss $\mathbf{A}$ is focused on generating realism and high-frequency details on the regions indicated by $\mathcal{M}$.~$\mathbf{A}$ not allows blurry outputs where the attention map $\mathcal{M}$ indicates, because all blurry regions are classified as fake and the adversarial loss tries to fix it by generating high-frequency details on these regions.

Finally, our full objective $\mathcal{L}$ is a mix of the reconstruction and the adversarial loss, maintaining the relevance of the reconstruction loss and scaling the adversarial loss by the hyperparameter $\lambda$.~Equation \ref{eq:fullloss} allows determining to what extent the adversarial loss $\mathbf{A}$ should influence to $\mathcal{L}$, thus, controlling the generation of artifacts in the output images.

\begin{equation}\label{eq:fullloss}
\begin{aligned} 
\mathcal{L}(G,D)&=\mathbf{R}(G) + \lambda\cdot\mathbf{A}(G,D).
\end{aligned}
\end{equation}

We perform ablation studies on the architecture, and verify the improvements of using our proposed attention mechanism.~Our ablation studies also consider the use and not of a pre-trained model in the generator.

\section{\uppercase{Experiments}}
\label{sec:experiments}

In this section, we describe the Flash and Ambient Illumination Dataset~(FAID) and the custom set of these images that we use.~We present the training protocol that we followed and show the quantitative and qualitative results that validate our proposal.~Finally, we present the controlled experiments that we perform to determine how the components of our architecture affect the overall performance.

\subsection{Dataset}

\begin{figure}[!h]
\centering
    \captionsetup[subfigure]{labelformat=empty}
    \begin{subfigure}[t]{0.153\textwidth}
      \centering
      \includegraphics[width=\linewidth]{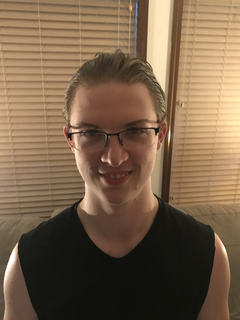}
    \end{subfigure}\hfill
    \begin{subfigure}[t]{0.153\textwidth}
      \centering
      \includegraphics[width=\linewidth]{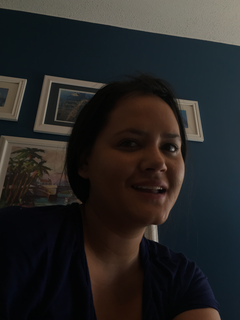}
    \end{subfigure}\hfill
    \begin{subfigure}[t]{0.153\textwidth}
      \centering
      \includegraphics[width=\linewidth]{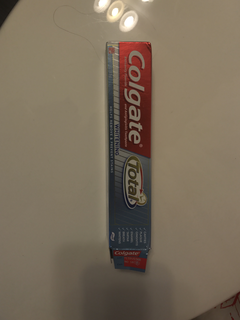}
    \end{subfigure}
    \vspace{8pt}
\caption{Ambient images from FAID~\cite{aksoy2018dataset} with low illumination, reflections, and shadows from external objects.}
\label{fig:dataset}
\end{figure}

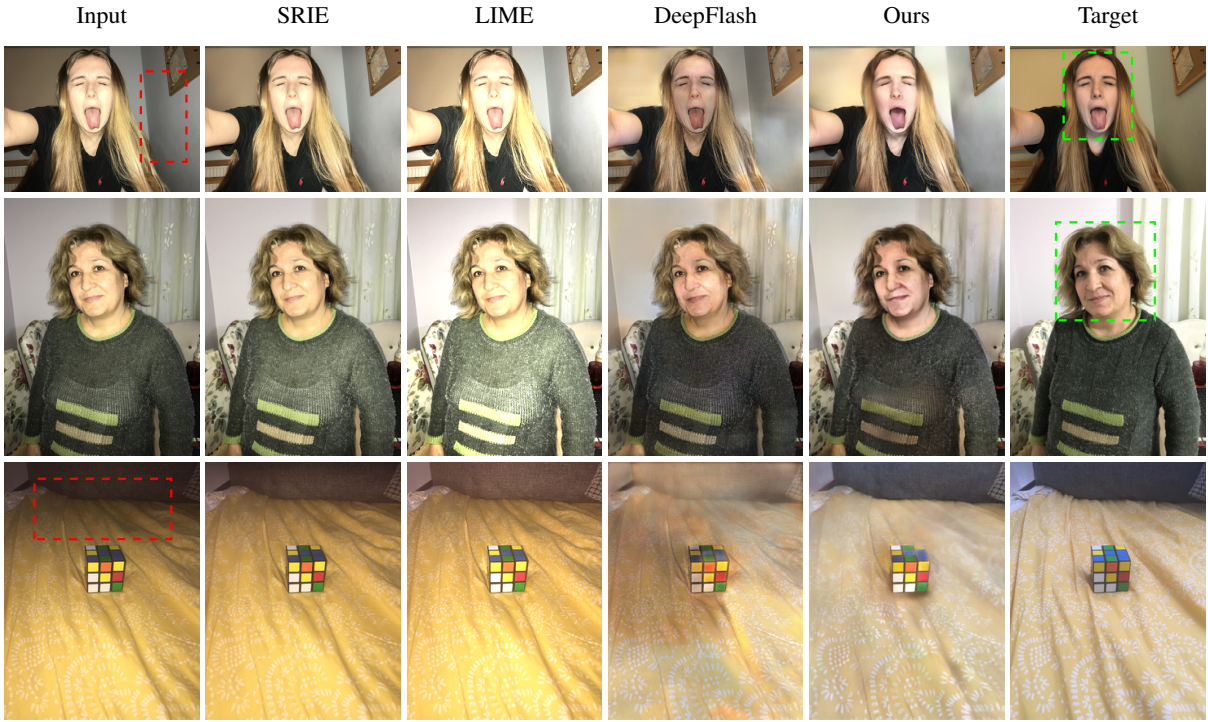
\begin{figure*}[!ht]
\centering
    \captionsetup[subfigure]{labelformat=empty}
    \begin{subfigure}[t]{0.162\textwidth}
      \centering
      \caption{Input}
      \begin{tikzpicture}
      \node[anchor=south west,inner sep=0] at (0,0) {\includegraphics[width=\linewidth]{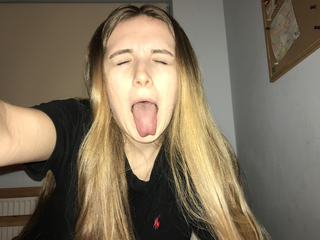}};
      \draw[red,dashed,line width=0.3mm] (1.8,0.4) rectangle (2.4,1.6);
      \end{tikzpicture}
    \end{subfigure}\hfill
    \begin{subfigure}[t]{0.162\textwidth}
      \centering
      \caption{SRIE}
      \includegraphics[width=\linewidth]{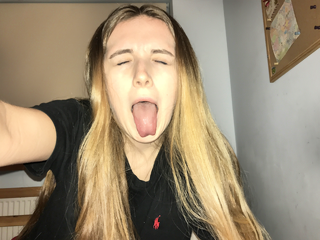}
    \end{subfigure}\hfill
    \begin{subfigure}[t]{0.162\textwidth}
      \centering
      \caption{LIME}
      \includegraphics[width=\linewidth]{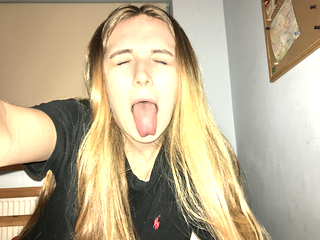}
    \end{subfigure}\hfill
    \begin{subfigure}[t]{0.162\textwidth}
      \centering
      \caption{DeepFlash}
      \includegraphics[width=\linewidth]{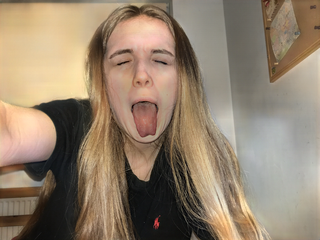}
    \end{subfigure}\hfill
    \begin{subfigure}[t]{0.162\textwidth}
      \centering
      \caption{Ours}
      \includegraphics[width=\linewidth]{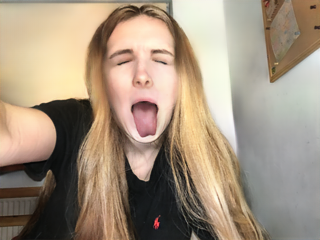}
    \end{subfigure}\hfill
    \begin{subfigure}[t]{0.162\textwidth}
      \centering
      \caption{Target}
      \begin{tikzpicture}
      \node[anchor=south west,inner sep=0] at (0,0){ \includegraphics[width=\linewidth]{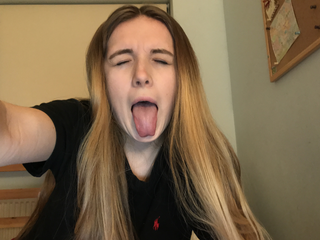}};
      \draw[green,dashed,line width=0.3mm] (0.7,0.7) rectangle (1.6,1.85);
      \end{tikzpicture}
    \end{subfigure}
    
    \vspace{1.5pt}
    \begin{subfigure}[t]{0.162\textwidth}
      \centering
      \includegraphics[width=\linewidth]{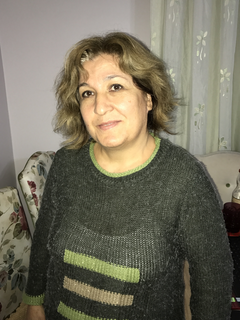}
    \end{subfigure}\hfill
    \begin{subfigure}[t]{0.162\textwidth}
      \centering
      \includegraphics[width=\linewidth]{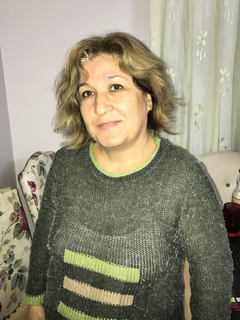}
    \end{subfigure}\hfill
    \begin{subfigure}[t]{0.162\textwidth}
      \centering
      \includegraphics[width=\linewidth]{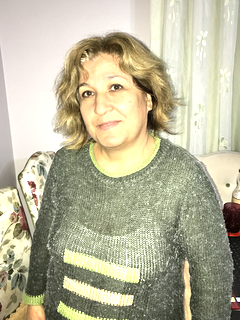}
    \end{subfigure}\hfill
    \begin{subfigure}[t]{0.162\textwidth}
      \centering
      \includegraphics[width=\linewidth]{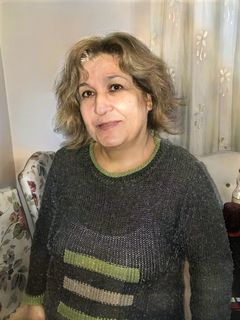}
    \end{subfigure}\hfill
    \begin{subfigure}[t]{0.162\textwidth}
      \centering
      \includegraphics[width=\linewidth]{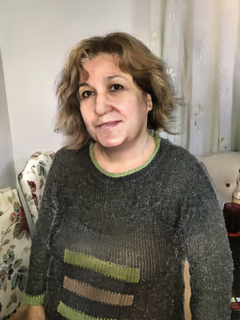}
    \end{subfigure}\hfill
    \begin{subfigure}[t]{0.162\textwidth}
      \centering
      \begin{tikzpicture}
      \node[anchor=south west,inner sep=0] at (0,0){
      \includegraphics[width=\linewidth]{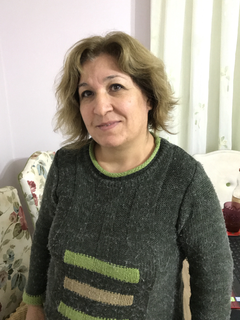}};
      \draw[green,dashed,line width=0.3mm] (0.6,1.8) rectangle (1.9,3.1);
      \end{tikzpicture}
    \end{subfigure} 

    \vspace{1.5pt}
    \begin{subfigure}[t]{0.162\textwidth}
      \centering
      \begin{tikzpicture}
      \node[anchor=south west,inner sep=0] at (0,0)
      {\includegraphics[width=\linewidth]{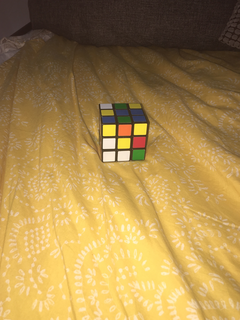}};
      \draw[red,dashed,line width=0.3mm] (0.4,2.4) rectangle (2.2,3.2);
      \end{tikzpicture}
    \end{subfigure}\hfill
    \begin{subfigure}[t]{0.162\textwidth}
      \centering
      \includegraphics[width=\linewidth]{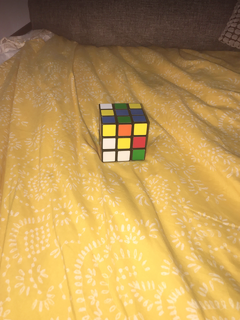}
    \end{subfigure}\hfill
    \begin{subfigure}[t]{0.162\textwidth}
      \centering
      \includegraphics[width=\linewidth]{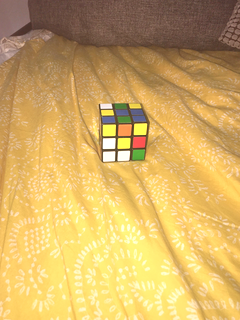}
    \end{subfigure}\hfill
    \begin{subfigure}[t]{0.162\textwidth}
      \centering
      \includegraphics[width=\linewidth]{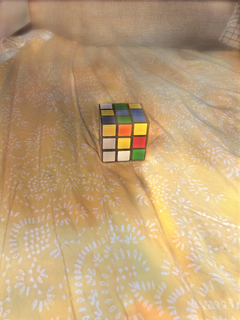}
    \end{subfigure}\hfill
    \begin{subfigure}[t]{0.162\textwidth}
      \centering
      \includegraphics[width=\linewidth]{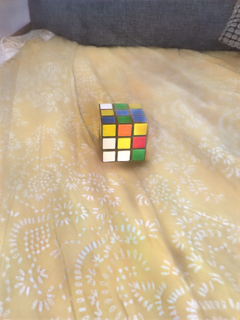}
    \end{subfigure}\hfill
    \begin{subfigure}[t]{0.162\textwidth}
      \centering
      \includegraphics[width=\linewidth]{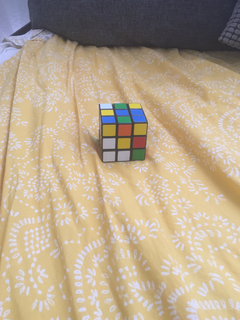}
    \end{subfigure} 
    
\caption{Qualitative comparison.~Enhancement of low-illuminated areas~(red), and estimation of natural skin and air tone of people~(green).~We compare with SRIE~\cite{Fu_2016_CVPR}, LIME~\cite{LIME}, and DeepFlash~\cite{CAPECE201928}.}
\label{fig:qualicomp1}
\end{figure*}

\noindent Introduced by \cite{aksoy2018dataset}, the FAID(Flash and Ambient Illumination Dataset) is a collection of pairs of flash and ambient images, which present 6 categories: \textit{People}, \textit{Shelves}, \textit{Plants}, \textit{Toys}, \textit{Rooms}, and \textit{Objects}.~As a result, we have $2775$ pairs of flash and ambient images.~We inspected each image in the dataset and found that there exist ambient images that have problems such as low illumination, shadows from external objects or even reflections.~Therefore, we used a reduced set of the entire FAID dataset for our experiments.~Finally, our custom dataset has $969$ pairs of images for training and 116 for testing and all images were resized to $320\times240$ or $240\times320$ depending on their orientation.

\subsection{Training}

\noindent We import all convolutional layers of the VGG-16 model~\cite{Simonyan15} in the encoder part of the generator, and train our model using the Adam optimizer~\cite{DBLP:journals/corr/KingmaB14} with $\beta_1=0.5$, based on~\cite{Isola_2017_CVPR}.~Using learning rates $2\cdot10^{-5}$ and $2\cdot10^{-6}$ for the generator and the discriminator respectively, equal or higher learning rate of the discriminator respect to the generator results on a divergence.~To regularize the adversarial loss $A$, we set $\lambda=1$, fewer values for $\lambda$ results on blurry outputs and higher values of $\lambda$ results on many artifacts.~The training procedure is performed using random crops of $224\times224$ and horizontal random flipping for data augmentation.~The implementation of our architecture is in Pytorch, and the training process takes approximately one day using an NVIDIA graphics card GeForce GTX 1070.

\subsection{Quantitative and Qualitative Validation}

\noindent We use the PSNR (Peak Signal-to-Noise Ratio) and the SSIM (Structural Similarity) to measure the performance of our quantitative results.~Table~\ref{Tab:psnrcomp} reports the mean PSNR and the mean SSIM on the test set, for 1000 epochs.~All hyperparameters are setting on the same way for~\cite{CAPECE201928}, and the encoder-decoder network was pre-trained on the ImageNet dataset~\cite{deng2009imagenet} instead on a model used for face recognition~\cite{Parkhi15}.~Our quantitative results do not significantly outperform the state-of-the-art image enhancement methods, but at least shows improvements on the flash image enhancement task.

\begin{table}[!ht]
\begin{center}
\begin{tabular}{l c c}
\toprule
 Method                         & PSNR  & SSIM \\
\midrule 
 LIME                           & 12.38 & 0.611 \\
 SRIE                           & 14.09 & 0.659 \\
 DeepFlash                      & 15.39 & 0.671 \\
 Ours                           & \textbf{15.67} & \textbf{0.684}  \\
\bottomrule
\end{tabular}
\caption{Reporting the mean PSNR and the mean SSIM with SRIE~\cite{Fu_2016_CVPR}, LIME~\cite{LIME}, and DeepFlash~\cite{CAPECE201928}.}
\label{Tab:psnrcomp}
\end{center}
\end{table}

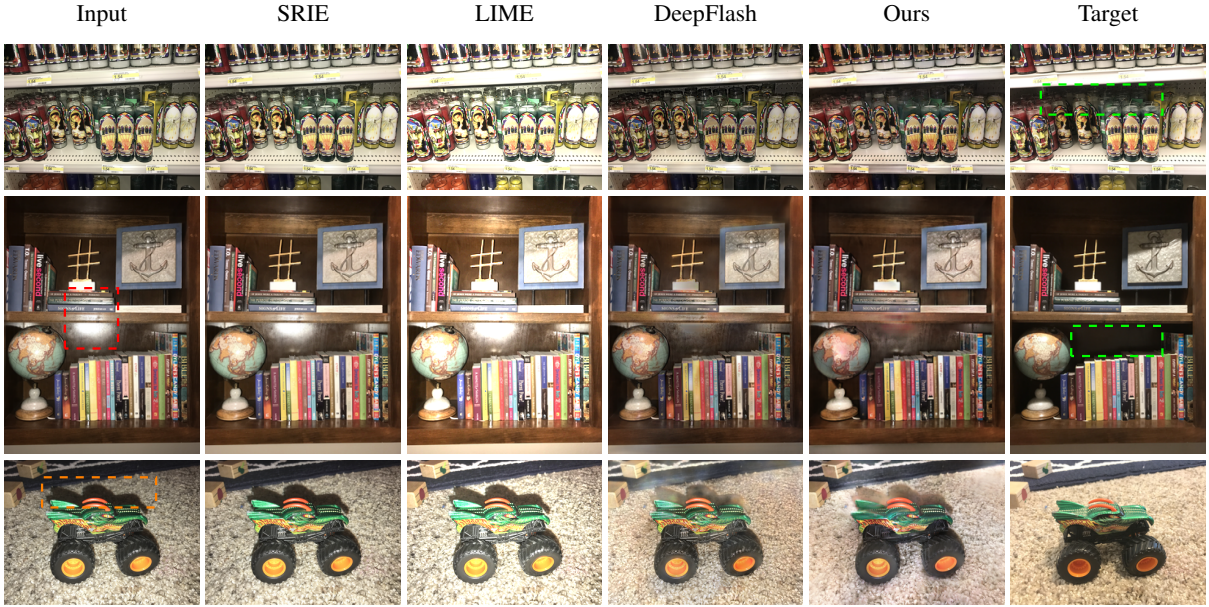
\begin{figure*}[!ht]
\centering
    \captionsetup[subfigure]{labelformat=empty}
    \begin{subfigure}[t]{0.162\textwidth}
      \centering
      \caption{Input}
      \includegraphics[width=\linewidth]{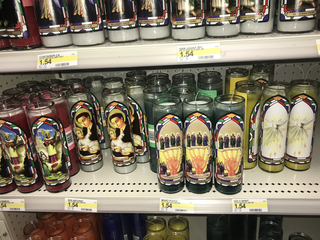}
    \end{subfigure}\hfill
    \begin{subfigure}[t]{0.162\textwidth}
      \centering
      \caption{SRIE}
      \includegraphics[width=\linewidth]{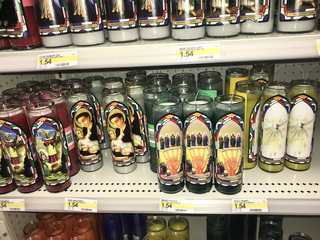}
    \end{subfigure}\hfill
    \begin{subfigure}[t]{0.162\textwidth}
      \centering
      \caption{LIME}
      \includegraphics[width=\linewidth]{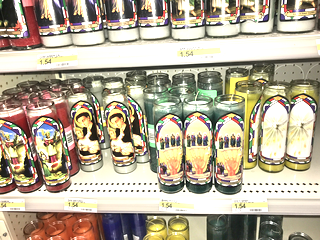}
    \end{subfigure}\hfill
    \begin{subfigure}[t]{0.162\textwidth}
      \centering
      \caption{DeepFlash}
      \includegraphics[width=\linewidth]{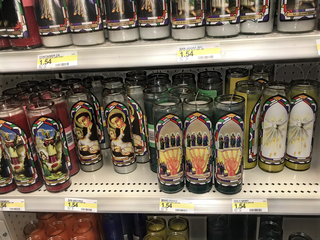}
    \end{subfigure}\hfill
    \begin{subfigure}[t]{0.162\textwidth}
      \centering
      \caption{Ours}
      \includegraphics[width=\linewidth]{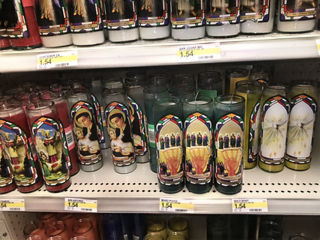}
    \end{subfigure}\hfill
    \begin{subfigure}[t]{0.162\textwidth}
      \centering
      \caption{Target}
      \begin{tikzpicture}
      \node[anchor=south west,inner sep=0] at (0,0){
      \includegraphics[width=\linewidth]{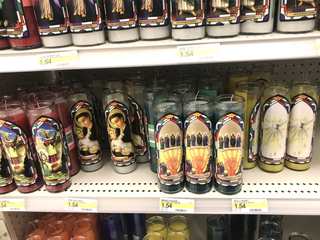}};
      \draw[green,dashed,line width=0.3mm] (0.4,1.0) rectangle (2.0,1.4);
      \end{tikzpicture}
    \end{subfigure}
    
    \vspace{1.5pt}
    \begin{subfigure}[t]{0.162\textwidth}
      \centering
      \begin{tikzpicture}
      \node[anchor=south west,inner sep=0] at (0,0) {\includegraphics[width=\linewidth]{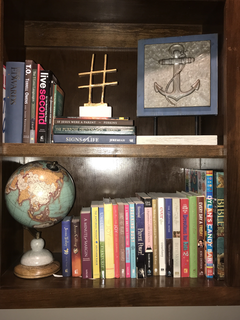}};
      \draw[red,dashed,line width=0.3mm] (0.8,1.4) rectangle (1.5,2.2);
      \end{tikzpicture}
    \end{subfigure}\hfill
    \begin{subfigure}[t]{0.162\textwidth}
      \centering
      \includegraphics[width=\linewidth]{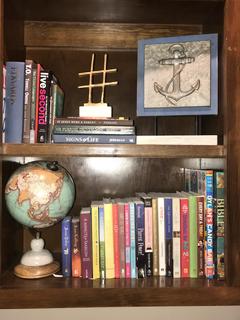}
    \end{subfigure}\hfill
    \begin{subfigure}[t]{0.162\textwidth}
      \centering
      \includegraphics[width=\linewidth]{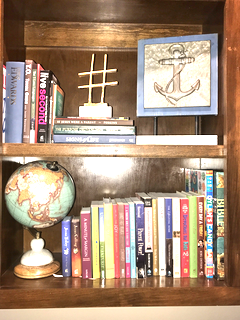}
    \end{subfigure}\hfill
    \begin{subfigure}[t]{0.162\textwidth}
      \centering
      \includegraphics[width=\linewidth]{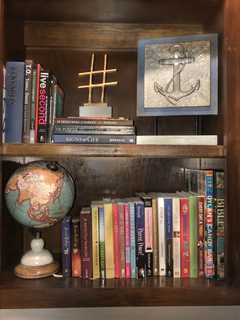}
    \end{subfigure}\hfill
    \begin{subfigure}[t]{0.162\textwidth}
      \centering
      \includegraphics[width=\linewidth]{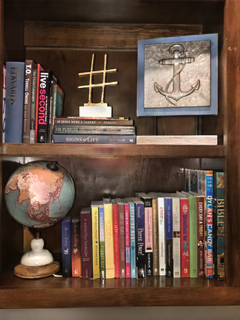}
    \end{subfigure}\hfill
    \begin{subfigure}[t]{0.162\textwidth}
      \centering
      \begin{tikzpicture}
      \node[anchor=south west,inner sep=0] at (0,0){ \includegraphics[width=\linewidth]{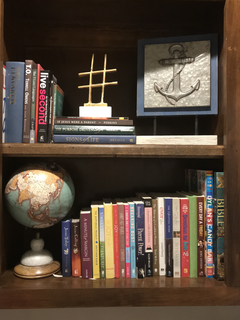}};
      \draw[green,dashed,line width=0.3mm] (0.8,1.3) rectangle (2.0,1.7);
      \end{tikzpicture}
    \end{subfigure}
    
    \vspace{1.5pt}
    \begin{subfigure}[t]{0.162\textwidth}
      \centering
      \begin{tikzpicture}
      \node[anchor=south west,inner sep=0] at (0,0) {
      \includegraphics[width=\linewidth]{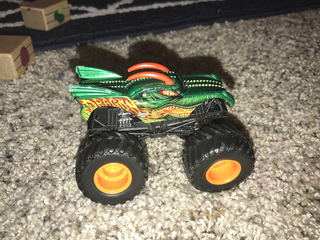}};
      \draw[orange,dashed,line width=0.3mm] (0.5,1.3) rectangle (2.0,1.7);
      \end{tikzpicture}
    \end{subfigure}\hfill
    \begin{subfigure}[t]{0.162\textwidth}
      \centering
      \includegraphics[width=\linewidth]{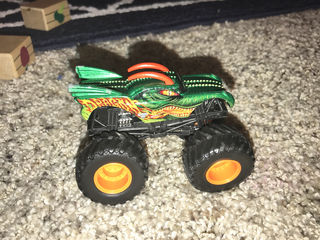}
    \end{subfigure}\hfill
    \begin{subfigure}[t]{0.162\textwidth}
      \centering
      \includegraphics[width=\linewidth]{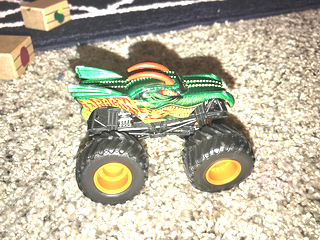}
    \end{subfigure}\hfill
    \begin{subfigure}[t]{0.162\textwidth}
      \centering
     \includegraphics[width=\linewidth]{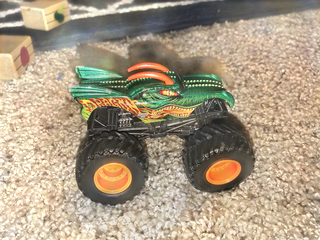}
    \end{subfigure}\hfill
    \begin{subfigure}[t]{0.162\textwidth}
      \centering
      \includegraphics[width=\linewidth]{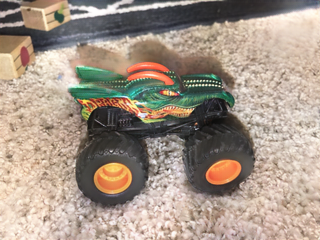}
    \end{subfigure}\hfill
    \begin{subfigure}[t]{0.162\textwidth}
      \centering
     \includegraphics[width=\linewidth]{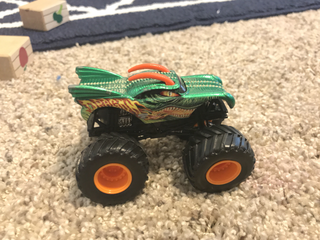}
    \end{subfigure} 
    
\caption{Qualitative comparison.~Generation of ambient shadows~(green), attenuation of overexposed areas~(red), and sideways shadow removal~(orange).~We compare with SRIE~\cite{Fu_2016_CVPR}, LIME~\cite{LIME}, and DeepFlash~\cite{CAPECE201928}.}
\label{fig:qualicomp2}
\end{figure*}

Estimation of the skin tone of people is shown in Figure~\ref{fig:qualicomp1}, where the illumination map created by LIME~\cite{LIME} conducts to brightening and overexposing the flash images.~LIME~\cite{LIME}, can not distinguish the natural color of dark objects and tend to illuminate them.~Results in SRIE~\cite{Fu_2016_CVPR} do not present considerable changes concerning the flash images on these kind of scenes.~DeepFlash~\cite{CAPECE201928} present non-uniform illumination on flash images of people, apparently this is due to trying to simulate shadows.~In the case of flash images that have low illuminated areas and also high illuminated areas like the Rubik's Cube on Figure~\ref{fig:qualicomp1}, \cite{CAPECE201928} present meaningless illumination on their results, and our method shows considerable better results, that is, our result looks much more similar to the ground truth.

Figure~\ref{fig:qualicomp1} reveals some aspects about the generation of ambient lighting on people.~Note the synthetic shadows in mouth and under the chin.~Almost all ambient images from train data was taken with light source that came from above through a typical light source that exists in homes.~Therefore, the model learns to generate synthetic ambient lighting simulating a light source that comes from above.

Figure~\ref{fig:qualicomp2} shows that our model synthesizes ambient shadows on flash images such as shelves, but suffer for restoring overexposed areas produced by the camera flash.~LIME~\cite{LIME}, and SRIE~\cite{Fu_2016_CVPR} do not attenuate overexposed areas or synthesize ambient shadows on these type of scenes, these methods do not handle this kind of issues of flash images.~DeepFlash architecture~\cite{CAPECE201928} performs weak ambient shadows, attenuate overexposed areas without restoring them, and outputs many artifacts on their results.~In the case of sideways shadow removal, all models fail~(including ours).

\subsection{Ablation Study}
\label{sec:controlledexp}

\begin{figure*}[!ht]
\centering
    \captionsetup[subfigure]{labelformat=empty}
    \begin{subfigure}[t]{0.163\textwidth}
      \centering
      \caption{Input}
      \includegraphics[width=\linewidth]{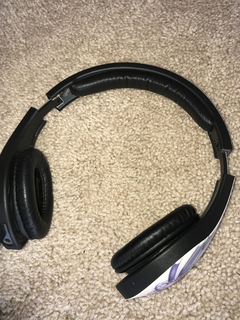}
    \end{subfigure}\hfill
    \begin{subfigure}[t]{0.163\textwidth}
      \centering
      \caption{Target}
      \includegraphics[width=\linewidth]{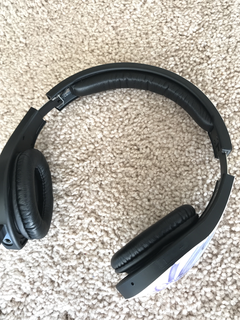}
    \end{subfigure}\hfill
    \begin{subfigure}[t]{0.163\textwidth}
      \centering
      \caption{\textbf{Default}}
      \includegraphics[width=\linewidth]{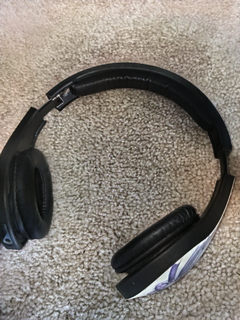}
    \end{subfigure}\hfill
    \begin{subfigure}[t]{0.163\textwidth}
      \centering
      \caption{$\mathbf{R}+\mathbf{A}$}
      \includegraphics[width=\linewidth]{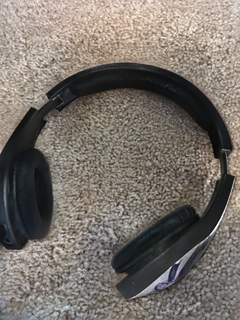}
    \end{subfigure}\hfill
    \begin{subfigure}[t]{0.163\textwidth}
      \centering
      \caption{$\mathbf{R}$}
      \includegraphics[width=\linewidth]{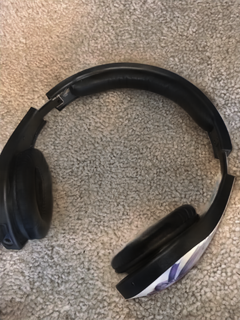}
    \end{subfigure}\hfill
    \begin{subfigure}[t]{0.163\textwidth}
      \centering
      \caption{U-Net}
      \includegraphics[width=\linewidth]{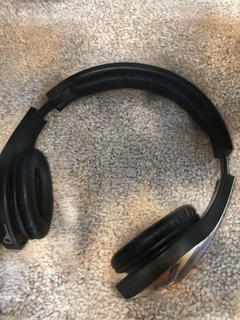}
    \end{subfigure}
    \vspace{1.5pt}
    \begin{subfigure}[t]{0.163\textwidth}
      \centering
      \includegraphics[width=\linewidth]{images/exp/comp/flash/Objects_148_flash.png}
    \end{subfigure}\hfill
    \begin{subfigure}[t]{0.163\textwidth}
      \centering
      \includegraphics[width=\linewidth]{images/exp/comp/ambient/Objects_148_ambient.png}
    \end{subfigure}\hfill
    \begin{subfigure}[t]{0.163\textwidth}
      \centering
      \includegraphics[width=\linewidth]{images/exp/comp/fake/Objects_148_flash.png}
    \end{subfigure}\hfill
    \begin{subfigure}[t]{0.163\textwidth}
      \centering
      \includegraphics[width=\linewidth]{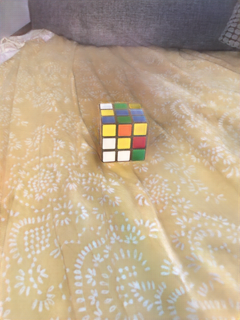}
    \end{subfigure}\hfill
    \begin{subfigure}[t]{0.163\textwidth}
      \centering
      \includegraphics[width=\linewidth]{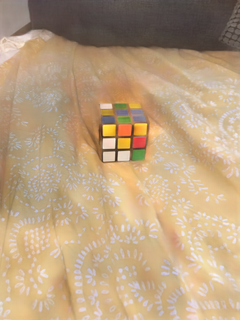}
    \end{subfigure}\hfill
    \begin{subfigure}[t]{0.163\textwidth}
      \centering
      \includegraphics[width=\linewidth]{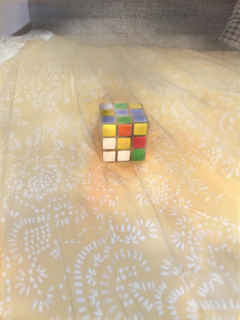}
    \end{subfigure}
\caption{Qualitative comparison for some configurations of the loss function, the use of our attention mechanism, and the network architecture.~\textit{Default} configuration represents the model that we use to compare with the state-of-the-art literature.}
\label{fig:controlledexp}
\end{figure*}

\noindent We perform different experiments to validate the final configuration of our architecture.~Table~\ref{Tab:contexp} reports the quantitative comparison between our controlled experiments.~Furthermore, we also show in Figure~\ref{fig:controlledexp} qualitative compositions between conditions in Table~\ref{Tab:contexp}.

\begin{table}[!ht]
\begin{center}
\begin{tabular}{l c c}
\toprule
  Condition                        & PSNR  & SSIM\\
\midrule
 1.~\textbf{Default~($\mathbf{R}_{\mathcal{M}}+\mathbf{A}_{\mathcal{M}}$)}                  & \textbf{15.67} & \textbf{0.684} \\  
 2.~$\mathbf{R}+\mathbf{A}$               & 15.55 & 0.676 \\
 3.~$\mathbf{R}$      & 15.64 & 0.681 \\
 4.~U-Net                          & 14.81 & 0.643 \\
\bottomrule
\end{tabular}
\caption{This table reports the mean PSNR and the mean SSIM for some configurations of the loss function, the use of our attention mechanism, and the network architecture.~\textit{Default} configuration represents the model that we use to compare with the state-of-the-art literature.}
\label{Tab:contexp}
\end{center}
\end{table}

Our quantitative assessments show that using a pre-trained model improves significantly the model trained from scratch~(condition 4).~The other methods seem to have similar results.~This is because these models, which use MAE for the objective function~(condition 3), generate blurry results to minimize the error between estimated images and the targets.~Condition 2, which is similar to the default model without the attention mechanism, has less quantitative values than condition 3 because the adversarial loss gives some sharpness on their output images.

We explore our qualitative results~(Figure~\ref{fig:controlledexp}) for different loss functions, the attention map, and network architectures.\\

\noindent \textbf{Loss function.}~Table~\ref{Tab:contexp} reports the influence by using the adversarial loss.~Condition 3 represents the same structure of the generator without considering the adversarial loss, i.e.,~just an encoder-decoder network, without a discriminator.~This architecture presents blurred results comparing with our default model.~In this case the reconstruction loss $\mathbf{R}$ is not enough to generate high-frequency details on their results, note the blurry image of the headphones~(Figure~\ref{fig:controlledexp}).~The adversarial loss $\mathbf{A}$ ensure a better quality due to the deep discriminator network, which classifies blurry results as fake.~Condition 2 presents also blurry results; however, the output images present more uniform illumination due to the adversarial loss.\\

\noindent \textbf{Attention map.}~Condition 1, which represent our default model, present uniform illumination, and high-frequency details~(note the sharpness on the headphone respect to the other conditions).~Our attention mechanism guides the reconstruction and adversarial loss to obtain uniform illumination and also sharpness results with less artifacts.~However, due to the robustness for overexposed areas and shadows, our model can not re-lighting dark areas with high-frequency details.~We believe that a better formulation of the attention mechanism could address this problem.\\

\noindent \textbf{Network architecture.}~As we report in Table~\ref{Tab:contexp}, we perform the well known U-Net~\cite{ronneberger2015u} architecture in condition 4.~We adopt the model proposed by \cite{Chen_2018_CVPR} for enhancing extreme low light images, and train it from scratch.~U-Net present blurry output images and also non-uniform illumination.~Our default model, which uses transfer learning, performs better quantitative and qualitative results.~We believe this is due to the few samples in the training set.

\section{\uppercase{Conclusions}}
\label{sec:conclusion}

\noindent Ambient lighting generation is a challenging problem, even more on flash images under low light conditions.~Shadows on the flash image have to be removed, overexposed areas should be reconstructed, and ambient shadows must be synthesized as a part of the simulation of an ambient light source.~In this paper, we propose a model with a guided reconstruction loss for normalizing the illumination and a guided adversarial loss to model high-frequency illumination details on flash images.~Our results show that our guided mechanism estimated high-frequency details without introducing visual artifacts in our synthetic ambient images.~The guided adversarial loss also produces more realistic ambient illumination on flash images than the state-of-the-art methods.~Our current results are promising, nonetheless, there are cases where our model fails such as: restoring overexposed areas, normalizing the lighting for flash images on extreme low light conditions, and sideways shadow removal on flash images~(see Figure~\ref{fig:qualicomp1}).~We believe that a more dedicated approach on the adversarial loss would be useful to address these issues.

Other methods based on intrinsic image decomposition~\cite{SHEN_TPAMI} would be also useful by recovering the albedo~(reflectance) and shading of the flash image, then, modifying directly the shading component to obtain the ambient image.~As we show on this article, some cases need a more dedicated treatment.~We aim to further study these cases and evaluate new techniques to improve the ambient lighting generation for flash images in such situations.

\section*{\uppercase{Acknowledgements}}

This work was supported by grant 234-2015-FONDECYT (Master Program) from Cienciactiva of the National Council for Science, Technology and Technological Innovation (CONCYTEC-PERU).~I thank all the people who directly or indirectly helped me with this work.


\bibliographystyle{apalike}
{\small
\bibliography{example}}



\end{document}